\documentclass[aps,prl,twocolumn]{revtex4}

\begin{document}

\title{``Disproof of Bell's Theorem" :  more  critics.}

\author{ Philippe Grangier}

\affiliation{Laboratoire Charles Fabry de l'Institut d'Optique, \\CNRS, Univ Paris-Sud,
CD128, 91127 Palaiseau, France}

\begin{abstract}
In a series of recent papers \cite{p1,p2,p3} Joy Christian claims to have ``disproved Bell's  theorem". Though his work is certainly intellectually stimulating, we argue below that his main claim is unwarranted. 
\end{abstract}

\maketitle

In a series of recent papers,  Joy Christian introduced a model which  violates Bell's inequalities, and which agrees with quantum mechanics, though at first sight it does look ``local  and realistic". 
For details the reader is referred to \cite{p1,p2,p3}, and here we will assume that the content of these papers is correct, and use the same notations. The essential features of the proposed model are thus : 

\begin{itemize}

\item the ``measurement results" denoted as $A_a(\mu) = \mu.a$ and $A_b(\mu) = \mu.b$ are algebraic quantities (bivectors) which do not commute, and which depend on the hidden variables ($\mu$, which is a trivector), the analysers directions ($a$ or $b$, which are unit vectors), and a sign $\pm 1$ which tells the outcome of the measurement, given $\mu$,  and either $a$ or $b$. 

\item  when averaged over $\mu$, the correlation functions deduced from these algebraic ``measurement results" are real numbers, which exactly reproduce quantum mechanical predictions, and thus violate Bell's inequalities. 

\end{itemize}

However, what is still lacking in that model  is a way to extract the ``sign" , ie the result of each individual (dichotomic) measurement, from the algebraic quantities $\mu.a$ or $\mu.b$.  As written in \cite{ p3}, we still need ``a yet to be discovered physical theory, which, when measured, should reproduce the binary outcomes $\pm 1$".

But whatever is this theory, it will give a real-valued function equal to $\pm 1$ for each particle and each measurement device, i.e. for each $\mu$, $a$ and $b$, and thus it will have to obey Bell's theorem... This brings us back to step 0 : nothing is wrong with Bell's theorem. 

In other words, though it seems that the only ``available" information in the algebraic quantities $\mu.a$ and $\mu.b$ are their signs, averaging over non-commuting algebraic quantities is certainly not equivalent to averaging over commuting real functions \cite{reply1}. So  the model proposed by Joy Christian  lacks an essential ingredient~: it does not provide a physical way to extract  the ``measurement results" from the algebraic quantities which are introduced. And whatever method will be used,  what has to be averaged according to  Bell's reasoning are the real-valued  functions (i.e. the measurement results), {\bf not } the algebraic quantities  \cite{reply2}. Then  Bell's theorem still tells us that the whole construction  must  be either non-local and/or non-realistic  \cite{note1}, or in conflict with quantum mechanics.

More generally, Bell's theorem cannot be ``disproved", in the sense that its conclusions follow from its premices in a mathematically correct way. On the other hand, one may argue that these premices are unduly restrictive, but this is discussing, not disproving \cite{pg}. Here the conclusion is that extracting the ``sign" of a bivector is a non-trivial operation~: this is an interesting mathematical remark, but not a challenge for Bell's theorem. 

In order to finally conclude this discussion, it may be enlightening to consider the following ``toy model"~: 

\begin{itemize}

\item The ``hidden variable" is a random variable $\epsilon$, with values $\pm 1$ with equal probabilities. If $\epsilon =1$, the particle goes along the Stern-Gerlach axis $a$, and if $\epsilon =-1$, it goes opposite to it.  For correlated particles 1 and 2 with $\epsilon_1 = - \epsilon_2 = \epsilon$ (singlet state), this model obeys Bell's inequalities, and  gives $S_{Bell}=2$.

\item But now let us change the nature of the ``measurement result", and consider that it is  the vector  $\epsilon_1 a$ for particle 1, and $\epsilon_2 b$ for particle 2.  In addition, let us define  the ``correlation function" for these two ``observables" \cite{reply2} by the scalar product $(\epsilon_1 a).(\epsilon_2 b)$. Since $\epsilon_1 \epsilon_2 = -1$ in all cases, the averaged correlation function is $-a.b$  alike quantum mechanics, and thus violates Bell's inequalities. According to the terminology of \cite{p1,p2,p3}, this is a ``local realistic model disproving Bell's inequalities". 

\end{itemize}

The crucial point in this toy model is  the following :  by computing correlation functions with just the same rules as  in \cite{p1,p2,p3}, the ``ordinary vectors" do just the same job as the ``sophisticated bivectors" : they violate Bell's inequalities. Therefore for achieving this goal  the  ``sophisticated bivectors" are completely useless. 

So once again : the basic problem in \cite{p1,p2,p3} is that  the mathematical objects which are used to calculate the  ``correlation functions" are simply not the good ones. According to Bell's formalism, which is based on usual classical statistics, these objects must be the measurement results $\pm 1$. But as soon as these objects are changed, and that new rules are introduced for computing  new  ``correlation functions",  one can easily violate Bell's inequalities, no matter whether the objects are ``ordinary vectors" or ``sophisticated bivectors".

Obviously, anybody may define new ways for calculating correlation functions if he wishes so, but this is not ``disproving Bell's theorem", because this moves too far out from Bell's hypothesis. What is at stake is rather an alternative formulation of Quantum Mechanics, and then the questions go to completely different grounds : is this new formulation correct ? is it useful ? can it handle more complicated situations such as multiparticle entanglement ? etc... Answering such questions is certainly more interesting than agitating useless polemics about an inexistent and irrelevant ``disproving". 

\vskip 2mm
\centerline{-o-o-o-} 
\vskip 2mm
Useful exchanges of ideas with Valerio Scarani and Gregor Weihs are acknowledged.


\begin{thebibliography}{1}

\bibitem{p1}
``Disproof of Bell's Theorem by Clifford Algebra Valued Local Variables", Joy Christian, arXiv:quant-ph/0703179

\bibitem{p2}
``Disproof of Bell's Theorem: Reply to Critics", Joy Christian, arXiv:quant-ph/0703244

\bibitem{p3}
``Disproof of Bell's Theorem: Further Consolidations", Joy Christian, arXiv:quant-ph/0707.1333

\bibitem{note1} Actually, by opposition to e.g. Bohm's model which is considered to be ``realistic but non-local", the present model can be said to be ``local but non-realistic", since one cannot associate simultaneous ``elements of reality"  to the two non-commuting measurements $a$ and $a'$ on one side  (or $b$ and $b'$ on the other side).   See e.g. in~\cite{p3} :``It is crucial to note here that a given bivector  $\mu . n$ cannot be spinning either ÒupÓ or ÒdownÓ, or in any other way, about any other direction but $n$ (...) Thus, our observables $A_a(\mu)$ and $B_b(\mu)$  represent quite faithfully what is actually observed in a Stern-Gerlach type experiment." This  is consistent with calculating  the correlation functions algebraically (like in quantum mechanics), and not from real-valued functions (which would lead to Bell's theorem). 


\bibitem{pg} An alternative view about \cite{p1,p2,p3} may be based on the ``contextual objectivity" point of view~\cite{pgp}~: by considering that the ``measurement result" is  the bivector $\mu.a$ rather than simply $\pm 1$, one makes the ``context" (i.e. the value of $a$) an intrinsic part of the measurement result, which certainly makes sense according to \cite{pgp}.  Then the algebraic ``hidden variable" $\mu$ can be seen as a way to carry the ``holistic" character of the entangled state, in a different way from the usual quantum formalism (usually $a$ and $b$ are seen as ``measurement parameters" rather than ``measurement results"). It is unclear whether or not such an approach might be interesting as an alternative formulation of quantum mechanics, but again  it does not  ``disprove"  Bell's theorem, and it contradicts local realism just as much as quantum mechanics does. 
\bibitem{pgp} ``Contextual objectivity and the quantum formalism",
Philippe Grangier, Int. J. Quant. Inf. {\bf 3:1}, 17 (2005); see also arXiv:quant-ph/0407025.

\bibitem{reply1} 
Besides the fact that \cite{p2} seems to ignore what a collegial tone is, it is constantly misinterpreting 
when accusing of misinterpretation. 
What is meant in this sentence  is the obvious fact that 
$$ \int (\mu.a)  (\mu.b) d\rho(\mu) \neq \int ``sign"(\mu.a)  ``sign"(\mu.b) d\rho(\mu), $$ where $``sign"$ is any function which gives the measurement result $\pm 1$, knowing $\mu.a$ and  $\mu.b$~: clearly the rhs has to obey Bell's theorem, while the lhs has not. 

\bibitem{reply2} 
It is actually quite revealing that  \cite{p2} keeps on using the wording ``observable" rather than ``measurement result". This ``quantum" vocabulary clearly misses  that the central issue in Bell's theorem, which is correlating clicks  between  detectors (corresponding to binary measurement results), and  not correlating  bivectors (which cannot be given any ``local realistic meaning"). More precisely, knowing the ``sign" of $(\mu.a)$ forbids to tell anything about  the ``sign" of $(\mu.a')$,  for the same  given $\mu$, while  in Bell's formalism $E(\lambda, a)$ and $E(\lambda, a')$ are two values of the same function, taken for the same $\lambda$ and two different measurement angles.  
So the proposed  model \cite{p1} cannot be a local realistic model, it could  at best be an alternative formulation of quantum mechanics \cite{note1}, like Bohm's theory is. 



\end{thebibliography}
\end{document}